%% file: main.tex
\documentclass[a4paper,11pt]{article}
\usepackage{pos}
\graphicspath{{figs/}}

\newcommand{\uhecrs}{\mbox{UHECRs}}

\newcommand{\offline}{\mbox{$\overline{\rm Off}$\hspace{.05em}\raisebox{.3ex}{$\underline{\rm line}$}}}

\title{Machine Learning for the EUSO-SPB2 Fluorescence Telescope Data Analysis}
\ShortTitle{Machine Learning for EUSO-SPB2}
\author*[a]{George Filippatos}
\affiliation[a]{Colorado School of Mines, Golden, USA}
\author[b]{Mikhail Zotov}
\emailAdd{gfilippatos@mines.edu}
\affiliation[b]{Skobeltsyn Institute of Nuclear Physics, Lomonosov Moscow State University\\
    Moscow, Russia}

\onbehalf{for the JEM-EUSO Collaboration\\[-1mm]{\normalsize \normalfont (a complete list of authors can be found at the end of the proceedings)}}

\abstract{%
The Extreme Universe Space Observatory on a Super Pressure Balloon 2 (EUSO-SPB2) is the most advanced balloon mission undertaken by the JEM-EUSO collaboration. EUSO-SPB2 is built on the experience of previous stratosphere missions, EUSO-Balloon and EUSO-SPB, and of the Mini-EUSO space mission currently active onboard the International Space Station.  EUSO-SPB2 is equipped with two instruments: a fluorescence telescope aimed at registering ultra-high energy cosmic rays (UHECRs) with an energy above 2~EeV and a Cherenkov telescope built to measure direct Cherenkov emission from cosmic rays with energies above 1~PeV. The EUSO-SPB2 mission will provide pioneering observations on the path towards a space-based multi-messenger observatory. As such, a special attention was paid to the development of triggers and other software aimed at comprehensive data analysis.  A whole number of methods based on machine learning (ML) and neural networks was developed during the construction of the experiment and a few others are under active development. Here we provide a brief review of the ML-based methods already implemented in the instrument and the ground software and report preliminary results on the ML-based reconstruction of UHECR parameters for the fluorescence telescope.
}

\ConferenceLogo{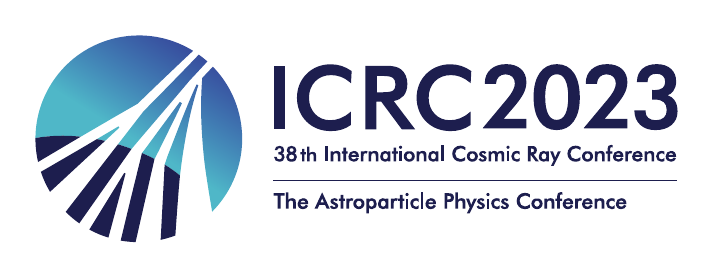}

\FullConference{
38th International Cosmic Ray Conference (ICRC2023)\\
  26 July - 3 August, 2023\\
  Nagoya, Japan
}

\begin{document}
\maketitle

\section{Introduction}

The Extreme Universe Space Observatory on a Super Pressure Balloon 2 (EUSO-SPB2) consisted of two optical telescopes, the Cherenkov telescope (CT) and the fluorescence telescope (FT). The CT was made up  of 512 silicon photo-multipliers and points towards the limb of the Earth, with the eventual goal of observing Earth-skimming $\nu_\tau$'s. The FT consisted of 108 multi-anode photo-multiplier tubes (MAPMTs) and pointed down at the atmosphere from the SPB's float altitude of 33 km. Designed to detect ultra high energy cosmic ray (UHECR) induced extensive air showers (EAS), the FT contained a total of 6,912 pixels in a rectangular grid of 48 x 144 pixels (12$^\text{o}$ x 36$^\text{o}$) each with an integration time of 1.05 $\mu$s. 

Serving as a stepping stone to future space based experiments, such as K-EUSO \citep{K-EUSO} or the Probe of Extreme multi-messenger Astrophysics \citep{POEMMA}, EUSO-SPB2 aimed to observe UHECRs under constraints similar to a satellite mission. This includes limited power and telemetry. Being powered by lithium-ion batteries charged from solar panels, and being designed to operate through extended nights if the balloon drifted south required that low power CPUs must be utilized. This imposes an intrinsic limit on the computational intensity of the onboard data handling processes. Additionally, limited telemetry means that only a fraction of recorded data can be downloaded. The planned primary telemetry connection for EUSO-SPB2, the Tracking and Data Relay Satellite System, would have allowed for only less than 10 percent of the data recorded with the FT to be downloaded during flight. To combat this limitation, SuperBIT, the other payload to launch during the 2023 Wanaka campaign, carried physical storage devices in parachute assemblies that were dropped over south America for data recovery. These limits on telemetry and power require sophisticated schemes for handling the large amount of data generated by the FT in flight. In the months immediately prior to launch, it was decided that a Starlink connection would be flown for the first time on a high altitude balloon. This connection allowed for roughly two orders of magnitude higher bandwidth allowing a significant fraction of the recorded data to be downloaded. 

Indirect measurements of UHECRs, such as those attempted by EUSO-SPB2, require some form of reconstruction in order to estimate the characteristics (energy, etc.) of the primary particle. With many changing variables, reconstructions of this nature pose a challenge. Degeneracy exists between the observed signals of EAS with different parameters. Additionally features of the detector complicate the analysis of signals. This includes gaps in between the different MAPMTs and non-uniformities in the response of the detector. By utilizing a neural network based approach to reconstruction of the EAS properties, these effects can be accounted for in an efficient manner. 

EUSO-SPB2 launched on May 13$^\text{th}$ from Wanaka New Zealand with hopes of a months-long flight. Unfortunately due to a leak in the balloon, the flight was terminated thirty seven hours later. Several hours of self triggered data were recorded and downloaded by the FT during this shortened flight. Given the accumulated exposure, the expected number of observed EAS was below one. Despite the shortened flight, the FT was able to record many thousands of laser shots mimicking the optical signature of an EAS during a field campaign in Delta Utah in fall 2022.  

\section{Data Sets}

\subsection{Simulations}

\begin{figure}[h!]
    \includegraphics[width=.48\textwidth]{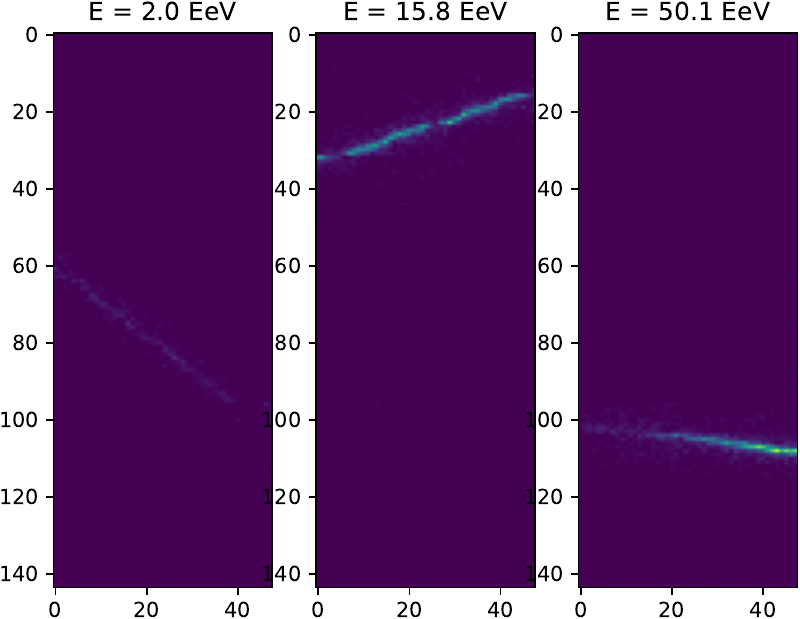}\quad\includegraphics[width=.48\textwidth]{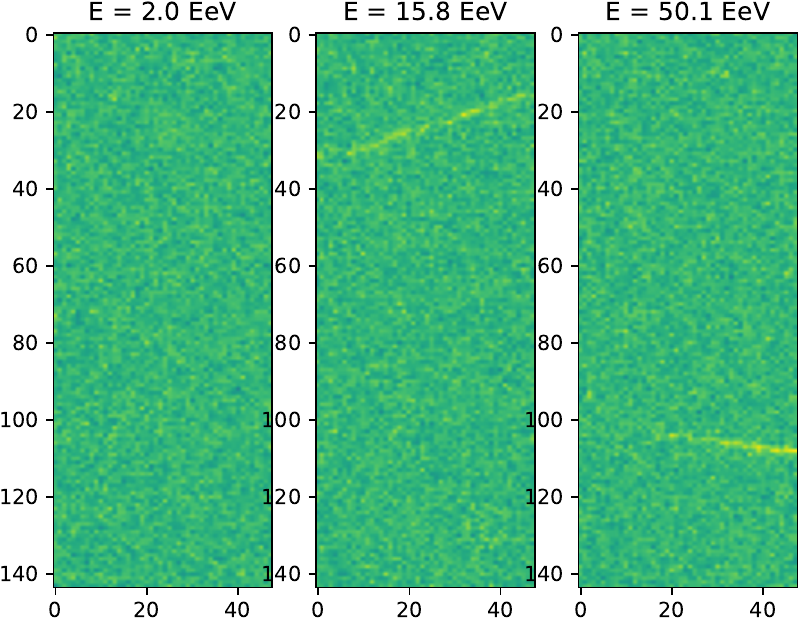}
    \caption{The left three panels: examples of integral tracks with zero background illumination for simulated EAS with energies 2.0~EeV, 15.8~EeV, and 50.1~EeV. The right three panels: integral signals during 128 GTUs for the same EASs with expected background illumination.}
    \label{fig:tracks}
\end{figure}

In order to test and develop analysis techniques, as well as inform the design of the instrument, extensive simulations of the detector response are preformed. These simulations are carried out in the JEM-EUSO \offline{}  framework \citep{JEMEUSOCollaborationPaulBertaina2017_1000082477}. Starting with a simulation of a shower profile in either Corsika or Conex, the fluorescence light from the EAS is propagated through the atmosphere to the detector, accounting for dispersion and scattering. The response of the detector is then simulated using a Geant4 implementation of the optics and electronics, which has been tuned based on laboratory measurements of the instruement. This creates a simulated output that mimics the recorded data: 6,912 traces 128 frames long. Examples of three simulated showers are shown in Figure \ref{fig:tracks}, both with and without simulated background. 

The result of the simulation can be used to test different configurations of the instrument, such as trigger parameters,and can be used for analysis such as estimating the energy resolution of the detector. One relevant aspect of these simulations is that there is significant degeneracy between the energy of the simulated EAS and the brightest signal recorded in the detector. This is the result of different geometries of the EAS. For example more inclined showers result in signal being localized in fewer pixels. Additionally, there are detector-specific features that complicate analyses like energy estimation such as the physical spacing between MAPMTs. These effects, along with the different geometries EAS may have, can be accounted for by utilizing a Monte Carlo approach and simulating many thousands of showers to be used as a training set. 

\subsection{Laser Events}

\begin{figure}[h!]
    \centering
    \includegraphics[width=.85\textwidth]{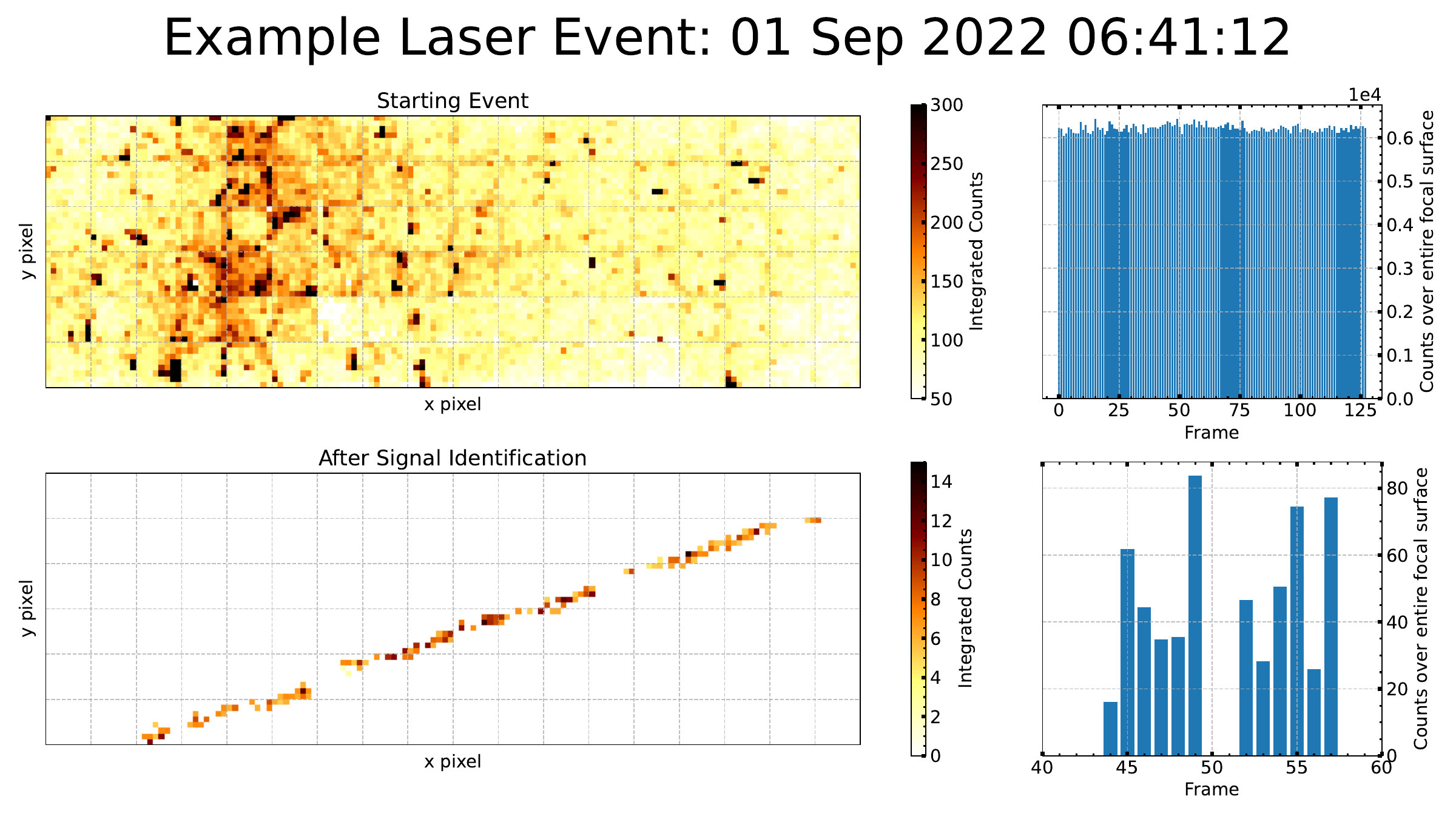}
    \caption{Example laser event recorded during field tests. Integrated signal over entire packet (top) and laser signal after offline identification (bottom). Energy of laser chosen to mimic a 3 EeV EAS, geometry chosen to cover entire focal surface with laser located 24 km away from the detecor.}
    \label{fig:laser}
\end{figure}

In addition to the simulated data sets, data taken on the ground during a field campaign prior to flight were available to develop and test ML methods. In the fall of 2022, the fluorescence telescope was transported to the Telescope Array Black Rock Mesa FD for calibrations and testing. A mobile laser trailer was parked 24~km away from the site in order to mimic the geometry expected of EAS during flight. This laser trailer is equipped with a steerable laser head, allowing for thge laser to be fired in any direction. It also includes an energy probe which samples and records every laser shot. The instrument was triggered internally providing an end to end verification of the data acquisition system. An example of a recorded laser shot is shown in Figure \ref{fig:laser}. Of note is the relatively low signal when compared to the entire recorded packet. 

\section{Binary Classification}

Due to limited telemetry onboard the super pressure balloon, an onboard scheme is needed to prioritize data for download. 
The chosen approach to this problem is to use a combination of recurrent and convolutional neural network.

A convolutional neural network (CNN) is a type of deep neural network that can be used as a binary classifier, and is most often used for image recognition.
There are several advantages to using a neural network approach to the problem of onboard data classification.
A major one is that the majority of the computationally intensive portion of the calculation needs to only be done once prior to the flight.
Once the model is trained, it can be efficiently used to classify data with minimal computational overhead.
Another major advantage of CNNs compared to other types of image classification algorithms is that they require minimal pre-processing.
Therefore no prior knowledge of what makes events recognizable is needed, as the model learns these features on its own.
Lastly, CNNs are shift and rotation invariant, meaning that the location of the signal in the camera is not relevant to the model's ability to classify it correctly.

Each frame of data recorded is passed through a convolutional layer. The output of this CNN is then fed into a recurrent neural network (RNN). This searches for signal moving in time as we expect the EAS induced signals to. Long short term memory gates were experimented with, but found to add no improvement to the performance of the classifier, while increasing the computational load. By utilizing just in time compilation, complex neural networks such as this can be utilized as onboard software running efficiently without a dedicated graphics processing unit. 

\begin{figure}[h!]
    \centering
    \includegraphics[width=.6\textwidth]{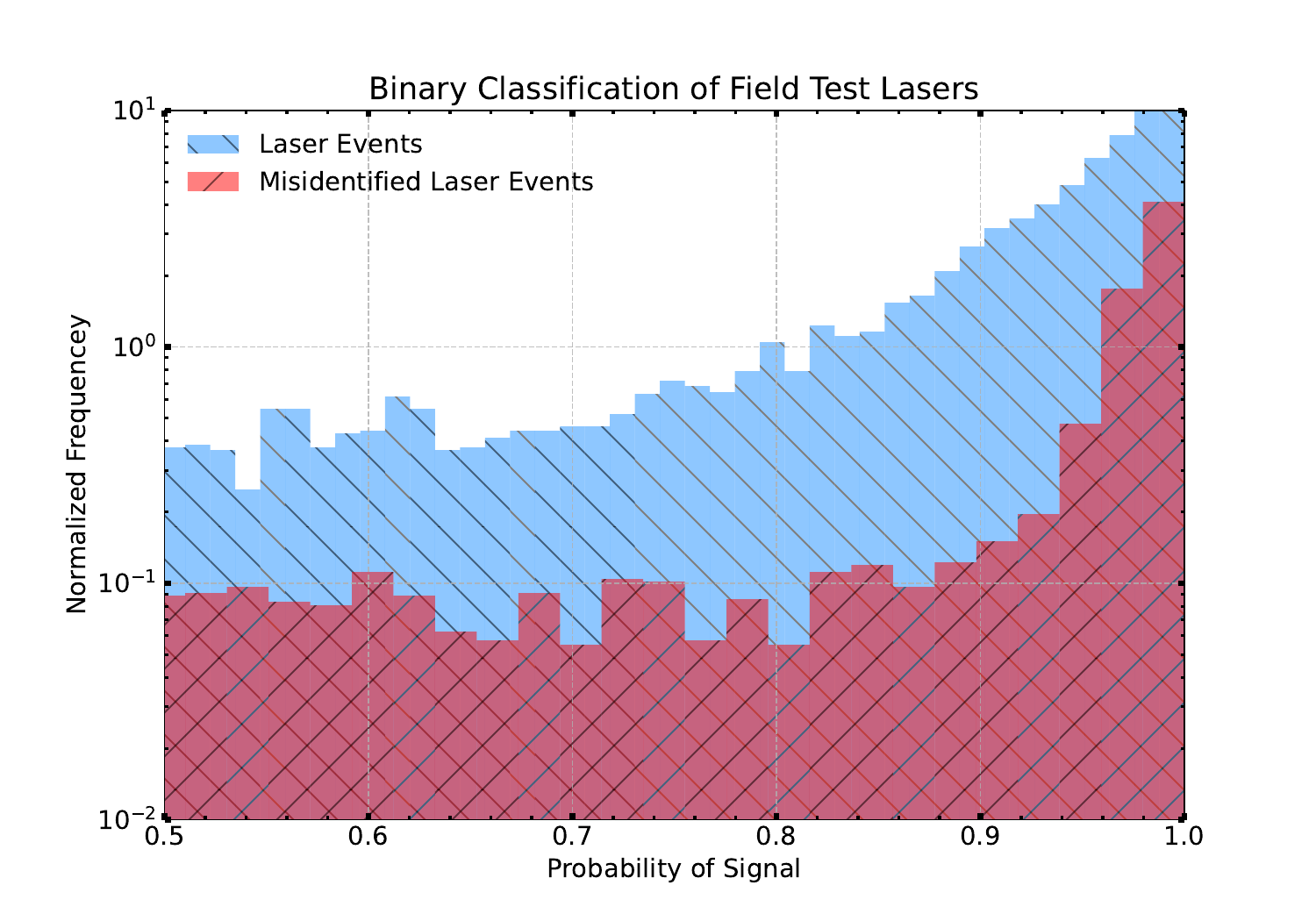}
    \caption{Distribution of probability assigned by onboard classifier to laser events (blue) and other triggered events (red).}
    \label{fig:laser_dist}
\end{figure}

The classifier was trained and developed on simulated data. On simulations, the model was able to reach near perfect accuracy. When applied to data outside of the training set, such as recorded data from previous EUSO missions, the fraction of events incorrectly identified as signal increased. The distribution of assigned probabilities for triggered laser events, and other triggered events is shown in Figure \ref{fig:laser_dist}. As can be seen, the laser events are correctly identified more than 90\% of the time, while other triggers are misidentified roughly 10\% of the time. 

\section{Energy Estimation}
One of the tasks of estimating initial parameters of a primary particle is reconstruction of its energy.  Application of machine learning methods to this problem has been studied for surface detectors of the Pierre Auger Observatory~\cite{2019ICRC...36..270G, 2021JInst..16P7019A} and the Telescope Array experiment~\cite{2020JPhCS1525a2001K, 2020arXiv200507117I}, see also~\cite{2018APh....97...46E}. However, as of time of this writing, we are not aware of any publications of these collaborations dedicated to the discussion of ML-based energy reconstruction of events registered with fluorescence telescopes.

On the other hand, this task has been attracting attention of experiments in gamma-astronomy for quite a while now~\cite{2008NIMPA.588..424A, 2017arXiv171106298L, 2019APh...105...44S, 2021arXiv210514927J, 2021arXiv210804130V, 2021arXiv210914262G, 2022ASPC..532..191N, 2022dlc2.confE...2G, 2022arXiv221203592I, 2023arXiv230211876A}. In this case, the main instruments are Cherenkov telescopes (Imaging Atmosphere Cherenkov Telescopes, IACTs), which register tracks produced by cascades initiated by gamma rays or by hadrons. Thus, the task of energy reconstruction for \uhecrs{} registered with a fluorescence telescope is similar to the corresponding task for Cherenkov telescopes. Two main approaches of energy reconstruction have been implemented for IACTs: the Random Forest method and various modifications of CNNs. CNNs have demonstrated better performance than Random Forest in most cases, thus we tried them as the first tool to test.

\begin{figure}[h!]
    \centering
    \includegraphics[width=.6\textwidth]{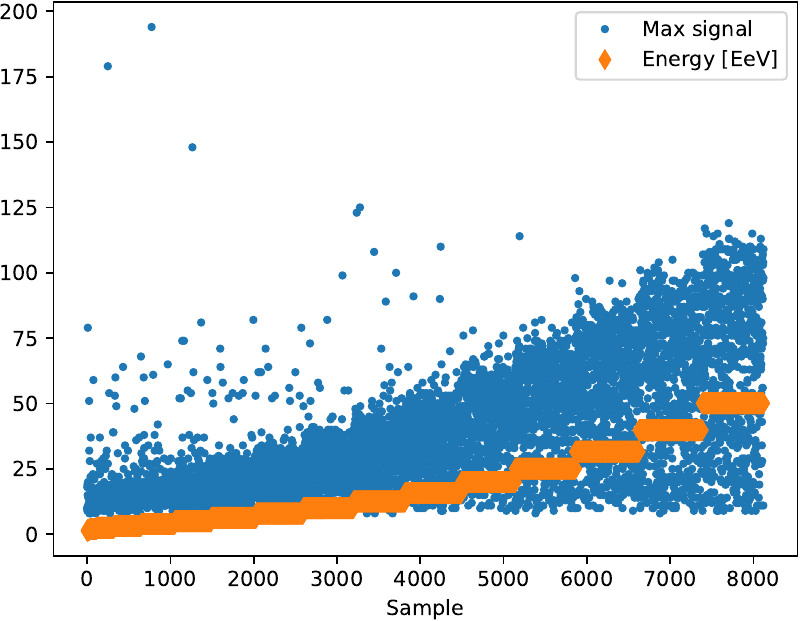}
    \caption{Energy (in EeV) and maximum signal amplitude (in photon counts)
    in the simulated training/testing data set.}
    \label{fig:e_signal}
\end{figure}

We took 8,123 events simulated with \offline~\cite{JEMEUSOCollaborationPaulBertaina2017_1000082477} as an input data set. The events have energies distributed uniformly wrt.\ the logarithm of energy in the range from $10^{18.1}$~eV (1.26~EeV) to $10^{19.7}$~eV (50.1~EeV), see Figure~\ref{fig:e_signal}, with azimuth angles covering the whole~$2\pi$ range and zenith angles varying from $0^\circ$ to $80^\circ$. All selected events intersected the field of view of the fluorescence telescope and registered a trigger. An example of three tracks arising from EAS initiated by cosmic rays with different energies are shown in Figure~\ref{fig:tracks}. Notice that tracks are rather dim for energies below $\sim10$~EeV.

We began with a simplified ``proof-of-concept'' problem setting. The initial data set contained integral signals with zero background illumination (like those shown in three left panels of Figure~\ref{fig:tracks}). We have tried several architectures of CNNs with different number of convolutional and fully connected layers and different hyper-parameters.  The model of choice consists of six convolutional layers and five fully connected layers though a few other models demonstrated similar performance. Adam with a varying learning rate was used as an optimization algorithm. Mean absolute percentage error (MAPE) was employed as a loss function (with mean squared logarithmic error showing similar results). In most cases, MAPE was below~10\%, which can be considered as a promising result.

\begin{figure}
    \includegraphics[width=.48\textwidth]{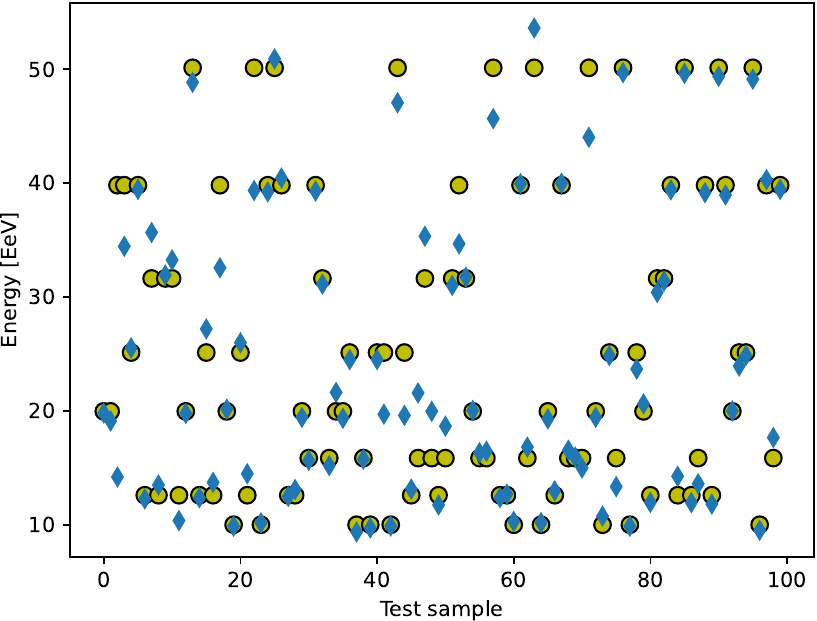}\quad\includegraphics[width=.48\textwidth]{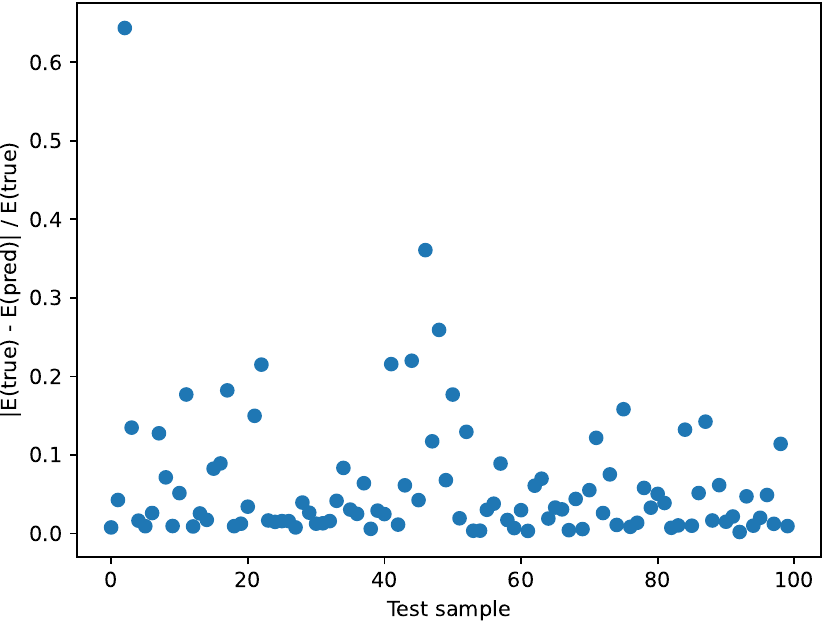}
    \caption{Left: an example of energy reconstruction for simulated EAS with energies $\ge10$~EeV with a CNN. True energies are shown by yellow circles, reconstructed energies are shown with blue diamonds. Right: relative errors of energy reconstruction for the same test sample.}
    \label{fig:ereco}
\end{figure}

Then we considered a more realistic case, i.e., signals with realistic intensity of background illumination (like those shown in the right three panels of Figure~\ref{fig:tracks}). We restricted our work to signals with $E\ge10$~EeV since signals with lower energies are too dim in a typical case. This left us with a data set comprised of 5,538 events. We applied a procedure of finding pixels that belong to a track similar to that developed for reconstruction of arrival directions. These reconstructed tracks were passed to the CNN, which demonstrated decent performance with MAPE being in the range 10--12\% in most runs. Figure~\ref{fig:ereco} shows an example of energy reconstruction for a testing data set. It can be seen that in most cases predicted energies are close to the true ones, with $\mathrm{MAPE}\approx7\%$.

We are working on studying richer simulated data sets and developing more sophisticated methods of energy reconstruction, as well as testing them on the data set of laser shots obtained during calibration of the instrument. These results will be reported in a dedicated paper.

\section{Summary}

EUSO-SPB2 was a pathfinder instrument which flew in May 2023. In addition to prototyping hardware for future experiments, EUSO-SPB2 offers an opportunity to develop analysis techniques for the next generation of space-based UHECR detectors. By utilizing extensive simulations, as well as ground observations of calibrated light sources, these analysis techniques can be qualified. We have shown that ML based approaches show promise as both onboard software, such as the binary classifier described here, and for determining characteristics of the primary particle such as the energy. While the flight of EUSO-SPB2 was shorter than expected, the analysis techniques developed show great promise for future experiments. 

\acknowledgments
The authors acknowledge the support by NASA awards 11-APRA-0058, 16-APROBES16-0023, 17-APRA17-0066, NNX17AJ82G, NNX13AH54G, 80NSSC18K0246, 80NSSC18K0473, 80NSSC19K0626, 80NSSC18K0464, 80NSSC22K1488, 80NSSC19K0627 and 80NSSC22K0426, the French space agency CNES, National Science Centre in Poland grant n. 2017/27/B/ST9/02162, and by ASI-INFN agreement n. 2021-8-HH.0 and its amendments. This research used resources of the US National Energy Research Scientific Computing Center (NERSC), the DOE Science User Facility operated under Contract No. DE-AC02-05CH11231. We acknowledge the NASA BPO and CSBF staffs for their extensive support. We also acknowledge the invaluable contributions of the administrative and technical staffs at our home institutions.
The research of MZ was funded by grant number 22-22-00367 of the Russian Science Foundation.

\bibliographystyle{JHEP-nt}
\bibliography{ml4spb2}
\input{JEM-EUSO_Authors_July2023_final_v2.tex}

\end{document}

%% file: JEM-EUSO_Authors_July2023_final_v2.tex
\newpage
{\Large\bf Full Authors list: The JEM-EUSO Collaboration\\}

\begin{sloppypar}
{\small \noindent
S.~Abe$^{ff}$, 
J.H.~Adams Jr.$^{ld}$, 
D.~Allard$^{cb}$,
P.~Alldredge$^{ld}$,
R.~Aloisio$^{ep}$,
L.~Anchordoqui$^{le}$,
A.~Anzalone$^{ed,eh}$, 
E.~Arnone$^{ek,el}$,
M.~Bagheri$^{lh}$,
B.~Baret$^{cb}$
% ...
}
\end{sloppypar}
\vspace*{.3cm}

{ \footnotesize
\noindent
% ...
$^{cb}$ Universit\'e de Paris, CNRS, AstroParticule et Cosmologie, F-75013 Paris, France\\
% ...
$^{ed}$ Istituto Nazionale di Fisica Nucleare - Sezione di Catania, Italy\\
% ...
$^{eh}$ INAF - Istituto di Astrofisica Spaziale e Fisica Cosmica di Palermo, Italy\\
% ...
$^{ek}$ Istituto Nazionale di Fisica Nucleare - Sezione di Torino, Italy\\
$^{el}$ Dipartimento di Fisica, Universit\`a di Torino, Italy\\
% ...
$^{ep}$ Gran Sasso Science Institute, L'Aquila, Italy\\
% ...
$^{ff}$ Nihon University Chiyoda, Tokyo, Japan\\ 
% ...
$^{ld}$ University of Alabama in Huntsville, Huntsville, AL, USA\\
$^{le}$ Lehman College, City University of New York (CUNY), NY, USA\\
% ...
$^{lh}$ Georgia Institute of Technology, USA\\
% ...
}